\documentclass[12pt]{article}

\usepackage{amsmath,amsfonts,bm}

\def\eqref#1{equation~\ref{#1}}
\def\1{\bm{1}}

\DeclareMathAlphabet{\mathsfit}{\encodingdefault}{\sfdefault}{m}{sl}
\SetMathAlphabet{\mathsfit}{bold}{\encodingdefault}{\sfdefault}{bx}{n}

\usepackage[left=2.5cm,
right=2.5cm,
top=2.3cm,
bottom=2.3cm,
headheight=20pt,
headsep=10pt,
footskip=25pt,
letterpaper]{geometry}
\usepackage{times}  % DO NOT CHANGE THIS
\usepackage{helvet}  % DO NOT CHANGE THIS
\usepackage{courier}  % DO NOT CHANGE THIS
\usepackage[hyphens]{url}  % DO NOT CHANGE THIS
\usepackage{graphicx} % DO NOT CHANGE THIS
\usepackage{natbib}  % DO NOT CHANGE THIS AND DO NOT ADD ANY OPTIONS TO IT
\usepackage{caption} % DO NOT CHANGE THIS AND DO NOT ADD ANY OPTIONS TO IT
\usepackage{algorithm}
\usepackage{algorithmic}
\usepackage[countmax]{subfloat}
\usepackage{booktabs}

\usepackage{tcolorbox}
\tcbuselibrary{breakable} % To allow boxes to break across pages
\usepackage{fancyvrb}    % For the Verbatim environment (capital 'V')
\tcbuselibrary{breakable,skins}   % “enhanced” 位于 skins 库中
\tcbuselibrary{listings}          % 可选：让 tcolorbox 里支持 verbatim/listing
\usepackage{enumitem}  % For customizing lists
\usepackage{fvextra}   % For line breaking in verbatim
\usepackage{amssymb}
\usepackage{multirow}
\DefineVerbatimEnvironment{Verbatim}{Verbatim}{breaklines=true, breakanywhere=true}

\definecolor{ysdarkpurple}{HTML}{4E2399}
\definecolor{ysshallowpurple}{HTML}{E6DBFF}
\definecolor{ysdarkred}{HTML}{8c2824}
\definecolor{ysshallowred}{HTML}{F8D7D7}
\definecolor{ysdarkblue}{HTML}{005E99}
\definecolor{ysshallowblue}{HTML}{CCEBFF}
\definecolor{ysdarkgrey}{HTML}{333333}
\definecolor{ysshallowgrey}{HTML}{E5E5E5}
\usepackage{tabularx, array}
\usepackage{ragged2e}
\usepackage{amsmath}
\usepackage{subcaption}
\newcolumntype{Y}{>{\centering\arraybackslash}X} % 居中可伸缩列

\definecolor{ColorGrok}{HTML}{FFFDE7}      % 保持不变
\definecolor{ColorPplx}{HTML}{EFFDFE}      % 保持不变
\definecolor{ColorOpenAI}{HTML}{F2F2F2}    % 将 EDEDED 进一步调浅为 F2F2F2 (非常浅的灰色)
\definecolor{ColorGemini}{HTML}{E6F4FE}    % 保持不变
\definecolor{ColorClaude}{HTML}{FFF3EB}    % 将 FFF0E5 进一步调浅为 FFF3EB (更极浅的桃色/米橙色)
\definecolor{SectionHeaderColor}{HTML}{FFFFFF} % 保持不变 (白色)

\colorlet{DarkerColorClaude}{ColorClaude!95!black}
\colorlet{DarkerColorPplx}{ColorPplx!95!black}
\colorlet{DarkerColorGemini}{ColorGemini!95!black}
\colorlet{DarkerColorOpenAI}{ColorOpenAI!95!black}
\colorlet{DarkerColorGrok}{ColorGrok!95!black}
\graphicspath{{figures/}}

\usepackage{makecell}
\usepackage{newfloat}
\usepackage{listings}
\DeclareCaptionStyle{ruled}{labelfont=normalfont,labelsep=colon,strut=off} % DO NOT CHANGE THIS
\floatstyle{ruled}
\newfloat{listing}{tb}{lst}{}
\floatname{listing}{Listing}

\usepackage{xcolor}
\definecolor{sentimentNegative}{HTML}{EA4335}
\definecolor{sentimentNeutral}{HTML}{4285F4}
\definecolor{mhKeywords}{HTML}{F4B400}
\definecolor{noMhKeywords}{HTML}{80868B}
\usepackage[utf8]{inputenc}
\usepackage[T1]{fontenc}
\usepackage[english]{babel}
\usepackage{amsmath,amsfonts,amssymb,thmtools,amsthm}
\usepackage{graphicx}
\usepackage{hyperref}
\usepackage{fancyhdr}
\usepackage{graphicx}
\usepackage{stfloats}
\usepackage{wrapfig}
\usepackage{epstopdf}
\usepackage{cleveref}
\usepackage{subfloat}
\usepackage{subcaption}
\usepackage{enumitem}
\usepackage{listings}
\usepackage{titlesec}
\usepackage{etoolbox}
\usepackage{setspace}
\usepackage{changepage}
\usepackage{etoolbox}
\usepackage{svg}
\usepackage[percent]{overpic}
\usepackage[colorinlistoftodos, shadow,color=blue!30!white
					,disable
]{todonotes}
\usepackage{xpatch}
\usepackage{siunitx}
\renewcommand{\sfdefault}{phv}

\renewcommand{\headrulewidth}{0pt}
\renewcommand{\footrulewidth}{0pt}
\fancypagestyle{first}{\fancyfoot[R]{\small\thepage}}

\setlist[itemize]{leftmargin=1em,itemsep=0ex,topsep=0ex}
\titlespacing*{\paragraph}{0pt}{0ex plus .1ex}{1ex}
\titlespacing*{\section}{0ex}{2.3ex plus .3ex minus .0ex}{.6ex plus .3ex minus .2ex}
\titlespacing*{\subsection}{0ex}{1.5ex plus .3ex minus .5ex}{.4ex plus .2ex minus .1ex}
\titlespacing*{\subsubsection}{0ex}{1.2ex plus .3ex minus .3ex}{.3ex plus .2ex minus .2ex}

\xapptocmd\normalsize{%
\abovedisplayskip=.8em plus .2em minus .2em
\belowdisplayskip=.6em plus .1em minus .1em
\abovedisplayshortskip=.8em plus .2em minus .2em
\belowdisplayshortskip=.6em plus .1em minus .1em
}{}{}

\setcitestyle{numbers}
\renewcommand{\cite}[1]{\citep{#1}}

\definecolor{mydarkblue}{rgb}{0.0,0.15,0.7}
\hypersetup{%
colorlinks=true,
linkcolor=mydarkblue,
citecolor=mydarkblue,
filecolor=mydarkblue,
urlcolor=mydarkblue}

\makeatletter
  \renewcommand{\maketitle}{%
    \begingroup
      {\centering\Large\@title\par}%
      \vskip 1em
      \centering
      \begin{tabular}[t]{@{}c@{}}\strut\@author\strut\end{tabular}%
      \vskip 0.3in minus 0.1in
    \endgroup
  }
\makeatother

\usepackage{xcolor}

\newcolumntype{L}{>{\RaggedRight\arraybackslash}X}

\usepackage{tablefootnote}

\title{Position: The Current AI Conference Model is Unsustainable! \\Diagnosing the Crisis of Centralized AI Conferences}

\date{}
\author{
Nuo Chen \quad
Moming Duan \quad
Andre Huikai Lin \quad
Qian Wang \quad
Jiaying Wu \quad
Bingsheng He \vspace{10pt} \\
National University of Singapore
}

\begin{document}
\pagestyle{fancy}

\maketitle
\thispagestyle{first}

% \footnotetext[1]{
% \hspace{\footnotesep} nuochen@comp.nus.edu.sg }
\begingroup
\renewcommand\thefootnote{}% 
\footnotetext{\hspace{\footnotesep} nuochen@comp.nus.edu.sg}
\endgroup

\begin{abstract}
Artificial Intelligence (AI) conferences are essential for advancing research, sharing knowledge, and fostering academic community. However, their rapid expansion has rendered the centralized conference model increasingly unsustainable. This paper offers a data-driven diagnosis of a structural crisis that threatens the foundational goals of scientific dissemination, equity, and community well-being. We identify four key areas of strain: (1) scientifically, with per-author publication rates more than doubling over the past decade to over 4.5 papers annually; (2) environmentally, with the carbon footprint of a single conference exceeding the daily emissions of its host city; (3) psychologically, with 71\% of online community discourse reflecting negative sentiment and 35\% referencing mental health concerns; and (4) logistically, with attendance at top conferences such as NeurIPS 2024 beginning to outpace venue capacity. These pressures point to a system that is misaligned with its core mission. In response, we propose the Community-Federated Conference (CFC) model, which separates peer review, presentation, and networking into globally coordinated but locally organized components, offering a more sustainable, inclusive, and resilient path forward for AI research.

\end{abstract}

\section{Introduction}
% \cn{arxiveffectcite:\cite{tran2020open}}
Academic conferences are the lifeblood of the Artificial Intelligence (AI) community. They serve as primary venues for disseminating cutting-edge research, facilitating invaluable networking, and shaping the trajectory of the field. The growth of major conferences like NeurIPS, ICML, and ICLR is a testament to the dynamism and impact of AI research.  Their core mission can be distilled into four pillars: (1) \textbf{The Scientific Mission}: Promoting AI research and enhancing academic exchange by serving as an efficient, peer-reviewed knowledge exchange platform. (2) \textbf{Knowledge Dissemination}: Sharing research results and recognizing thought leaders through presentations and awards. (3) \textbf{Community Building}: Fostering collaborations and a sense of belonging among researchers. (4) \textbf{The Social Contract}: Promoting diversity, equity, and inclusion (DEI) through inclusive practices.
However, this success has come at a cost. The prevailing model (a handful of massive, centralized, in-person events) is straining under its own weight~\cite{yang2025position,kim2025position}, raising a critical question: Is the current AI conference model sustainable?

\begin{figure}[t]
    \centering
    \includegraphics[width=0.69\linewidth]{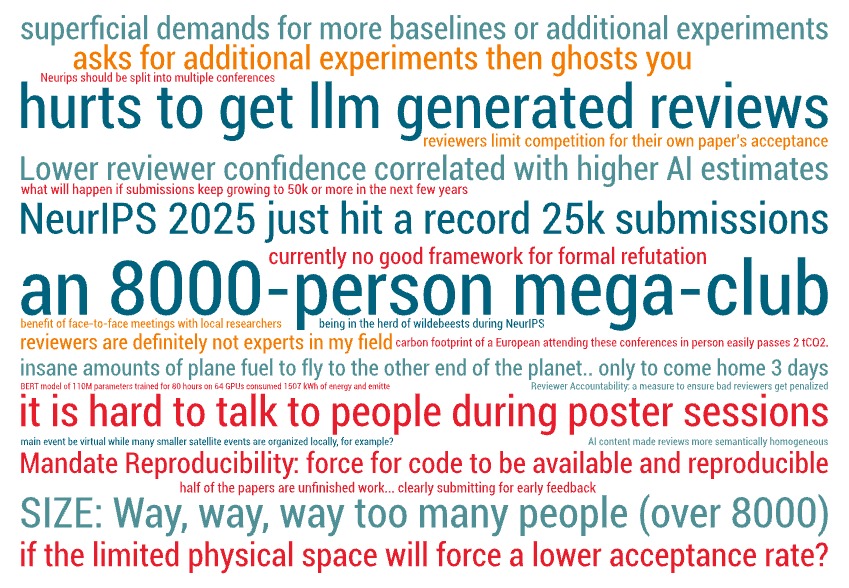}
    \caption{Sentence cloud of community grievances from online discussions.}
    \label{fig:wordcloud}
    \vspace{-2mm}
\end{figure}

The challenges are multifaceted: (1) The rapid growth in paper submissions (see Figure~\ref{fig:pubcount}) has overwhelmed the high-pressured peer-review system, often compromising review quality and fostering a culture where constructive criticism is discouraged and creates a hyper-competitive atmosphere that can stifle creativity; (2) Environmentally, the carbon footprint (see Figure~\ref{fig:carbon_pair}) generated by thousands of researchers traveling across the globe is significant and increasingly at odds with global sustainability goals \cite{Jackle2022carbon,Gokus2019climate}; (3) The high-stakes environment exacts a human toll. Researchers, especially doctoral students, face immense pressure (see Figure~\ref{fig:wordcloud}) to publish in these top-tier venues, leading to widespread anxiety, job insecurity, and a diminished sense of academic belonging \cite{nicholls2022wellbeing}; (4) The sheer scale of these events not only makes meaningful connections difficult, ironically isolating individuals within a crowd of thousands, but also creates venue capacity bottlenecks (see Figure~\ref{fig:nips_venue_predict}). 
% The sheer scale of these events can also make meaningful connection difficult, ironically isolating individuals within a crowd of thousands.

\begin{table*}[t]
\centering
\footnotesize
\renewcommand{\arraystretch}{1.6}
    \newcommand{\goalSA}{\textcolor{blue}{$\bullet$}}
    \newcommand{\goalKD}{\textcolor{orange}{$\blacksquare$}}
    \newcommand{\goalCB}{\textcolor{green}{$\blacktriangle$}}
    \newcommand{\goalSC}{\textcolor{red}{$\blacklozenge$}}
\resizebox{\textwidth}{!}{
\begin{tabular}{ 
>{\raggedright\arraybackslash}p{0.6cm}| 
>{\raggedright\arraybackslash}p{1.8cm}| 
>{\raggedright\arraybackslash}p{5.3cm}| 
>{\raggedright\arraybackslash}p{3.9cm}| >{\raggedright\arraybackslash}p{4.8cm}@{}
}
\hline
\textbf{Rel.} & \textbf{Metric / Indicator} & \textbf{Key Finding} & \textbf{Challenge} & \textbf{Community-Federated Conference (CFC) Solution} \\
\hline

\goalSA \goalKD & Per-author publication rate & The average annual publication rate in the AI field exceeds 4.5 papers, doubling in ten years, leading to a focus on quantity over quality. & Hyper-competition, SOTA hacking, or incremental contributions without conceptual depth. & Add submission windows to improve manuscript quality through rolling acceptance. \\
\hline

\goalSA \goalCB \goalSC  & Faculty average contribution  & The average contribution is growing exponentially, projected to exceed one paper every 2 months by 2030. &  Meaningful exchanges are diminished by fostering a high-pressure environment. & Decoupling submission alleviates deadline pressure and encourages deeper, more thoughtful research. \\
\hline

\goalSC \goalCB & Carbon footprint \& Travel barriers & NeurIPS 2024's travel emissions alone ($>$8,254 tCO$_2$e) exceed the daily emissions of the city of Vancouver. & Environmentally unsustainable, and create economic and visa barriers that undermine equity. & More than 90\% of attendees come from regional hubs in the region, significantly reducing emissions, travel, and costs. \\
\hline

\goalCB \goalSC \goalSA & Online negativity \& Mental health strain & Of the 405 threads on the reddit, $>$71\% of conference-related posts are negative, and 35\% contain mental health keywords. & High-pressure environ -ments erode community trust and inhibit innovative risk-taking. & Regional hubs of 50-200 people foster stronger peer connections and a sense of belonging. \\
\hline

\goalSA \goalKD & Conference Statistics \& Research lifecycle lag & Rejections grow exponentially, while the AI research lifecycle outpaces conference cycles, often rendering results outdated before presentation. & Resubmission strains reviewer capacity. Knowledge exchange is delayed and stale. & Rolling peer review cycles and multiple lightweight hubs enable more frequent dissemination of results. \\
\hline

\goalSC \goalCB \goalKD & Venue capacity bottleneck & Attendance at top AI conferences, exemplified by NeurIPS 2024, is beginning to outpace venue capacity. & Exclusivity (inability to accommodate everyone) reduces the effectiveness of knowledge dissemination and community building. & Federated hubs, combined with a strong digital layer, ensure broad and equitable participation regardless of physical capacity. \\
\hline
\end{tabular}
}
\caption{Overview of conference challenges and solutions, illustrating the many-to-many \textit{rel}ationship between issues and core goals: \goalSA{} Scientific Advancement, \goalKD{} Knowledge Dissemination, \goalCB{} Community Building, \goalSC{} Social Contract: DEI$^{2}$.}
\label{tab:conference_overview_symbols_revised}
\end{table*}
\footnotetext[2]{In this paper, DEI principles specifically refer to ensuring a diverse, equitable, and inclusive participation environment in academic conferences. For a detailed discussion, see reference \cite{tammy2024aiconf}.}

In this position paper, we argue that the current centralized AI conference model is increasingly unsustainable, as it undermines the core goals of such conferences. Drawing from prior analyses \cite{tammy2024aiconf,zhang2022enhancing}, we identify four primary goals detailed in Table \ref{tab:conference_overview_symbols_revised}.

These goals are being challenged by unsustainable growth. Our findings reveal: For the scientific mission, the shrinking lifecycle of AI research (e.g., often $<$7 months) clashes with annual cycles, leading to outdated presentations. Knowledge dissemination is eroded by hyper-competition, SOTA hacking, and p-hacking, creating an illusion of progress. Community building suffers from dispersed citation networks and heightened anxiety, as evidenced by negative sentiments in online discussions. These challenges are compounded by high carbon emissions (e.g., NeurIPS 2024’s estimated 8,254 tCO$_2$e from travel), venue limits, and mental health strains, all of which violate the social contract and exacerbate inequities.

To address these, we propose the Community-Federated Conference (CFC) model, a decentralized framework that decouples peer review, dissemination, and networking into interconnected layers: unified global peer-review with rolling cycles, federated regional hubs for localized interactions, and a digital layer for global connectivity. This model mitigates the identified challenges, aligning better with the four core goals. We conclude with a call for the AI community to embrace this shift toward sustainability and equity.

This paper provides a data-driven diagnosis of this unsustainability. We first establish our methodology for assessing the crisis (Section~\ref{sec:methodology}). We then show how the model fails across the four core conference goals (Section~\ref{sec:badissue}). After arguing that simple, incremental fixes are insufficient (Section~\ref{sec:fix}), we conclude with a call for the AI community to embrace a paradigm shift toward sustainability and equity.

\vspace{-2mm}

\section{Goals and Core Elements of AI Conferences}
\label{sec:goals}

Artificial intelligence (AI) academic conferences (e.g., NeurIPS \cite{neurips}, ICLR \cite{iclr}, and ICML \cite{icml}) have evolved from mere platforms for the release of technical achievements into multifunctional ecosystems with far-reaching influence, undertaking multiple missions such as promoting industry development, interdisciplinary integration, and social responsibility. In this section, we analyze the goals and core elements of AI conferences as follows:

\paragraph{Scientific Mission: Promoting AI Research and Enhancing Academic Exchange}
% \cn{how to define useful exchange}
% 促进人工智能研究，加强学术交流
The most basic and core goal of AI conferences is to serve as \textbf{an efficient, peer-reviewed knowledge exchange platform} \cite{zhang2022enhancing} to promote AI research \cite{tammy2024aiconf,ZHANG2021studyonai}. For instance, ICML explicitly calls for papers presenting          ``significant, original, and previously unpublished research'' in all areas of machine learning \cite{icml}, highlighting its commitment to advancing the scientific frontier. Similarly, ICLR highlights its interest in ``cutting-edge research on all aspects of deep learning, representation learning, and related areas''\cite{iclr}, positioning itself as a primary venue for presenting state-of-the-art discoveries. Being accepted by a top AI conference means that a research has been recognized by the community and valued its exchange for advancing the AI domain.

\paragraph{Knowledge Dissemination: Sharing Research Results and Recognizing Thought Leaders}
Beyond promoting research, AI conferences act as essential venues for the dissemination of state-of-the-art results and the recognition of influential contributors in the field. As an example, Direct preference optimization \cite{rafailov2023dpo} won NeurIPS 2023 outstanding paper runner-up for its direct preference optimization method, and Adam \cite{kingma2017adam} received test of time award due to its lasting contributions to optimization in machine learning. The presentation and discussion of the accepted and honored research motivates researchers to achieve broader impact.

\paragraph{Community Building: Sparking New Ideas and Collaborations, Exploring AI’s Future}

AI conferences foster community building by stimulating innovative ideas, encouraging collaborations, and exploring future directions in the field, with NeurIPS rules encouraging ``workshops to provide an informal, dynamic venue for discussion of work in progress and future directions'' \cite{nips_workshop}, an ICML workshop discription supporting ``researchers to share their latest results and ideas'' \cite{icml_workshop}, and ICLR explicitly stating in its mission that ``it is the premier gathering of professionals dedicated to the advancement of the branch of artificial intelligence'' \cite{iclr}. These elements, such as topic panels and networking events, spark new partnerships and accelerate the evolution of AI.

\paragraph{Social Contract: Promoting Diversity, Equity, and Inclusion}

In recent years, AI conferences have increasingly recognized their “social contract” by embedding the principles of diversity, equity, and inclusion (DEI) into their core mission. As argued by \cite{tammy2024aiconf}, there is a critical need to reimagine AI conference mission statements to explicitly promote inclusion. This involves concrete actions such as offering travel grants to underrepresented groups, establishing mentorship programs, enforcing strict codes of conduct. By actively fostering a more diverse and equitable community, conferences aim to mitigate biases in AI systems and ensure that the future of artificial intelligence is shaped by a wider range of perspectives, ultimately better serving all of humanity.

\section{Methodology for Assessing the Conference Crisis}
\label{sec:methodology}

To provide a rigorous, evidence-based critique of the current AI conference model, we developed a multi-pronged methodology combining quantitative analysis of publication trends, environmental impact modeling, and qualitative sentiment analysis. This section outlines our data sources and analytical framework.

\paragraph{Data Sources} Our analysis is grounded in a diverse set of public data sources:
\begin{itemize}
\item \textbf{CSRankings.org}: A publicly available dataset \cite{csranking_github} (2015-2024) was used to track publication volumes and faculty affiliations.
\item \textbf{Conference Statistics}: Official data from conference websites and fact sheets, particularly from NeurIPS \cite{nips_factsheet}, provided submission, acceptance, and attendee numbers.
\item \textbf{Social Media Corpus}: A collection of 405 discussion threads was programmatically gathered from the r/MachineLearning subreddit.
\item \textbf{Author Affiliation Data}: Institutional affiliations of first authors were extracted from accepted papers to model global travel patterns.
\end{itemize}

\paragraph{Analytical Framework} The following analytical techniques were employed to quantify the challenges detailed in the subsequent section:
\begin{itemize}
\item \textbf{Publication and Productivity Metrics}: Analysis of CSRankings data involved tracking total publication volumes and calculating the \textit{Per-capita Contribution}. This latter metric was derived by normalizing publication counts against the number of \textit{Effective Faculties} (an approach adapted from \cite{iclrpoint}) to measure academic productivity pressure.

\item \textbf{Environmental Impact Modeling}: The carbon footprint of conference travel was modeled by multiplying activity data (estimated round-trip flight distances from author affiliations) with established emission factors \cite{Gokus2019climate, defraHospitality} to yield a CO$_2$ equivalent (tCO$_2$e) estimate for events.

\item \textbf{Community Sentiment Analysis}: The Reddit corpus was subjected to computational analysis to assess community well-being. This process included applying the VADER sentiment tool to determine overall sentiment polarity and performing keyword frequency analysis to measure the prevalence of mental health indicators like ``anxiety'' and ``burnout''.

\item \textbf{Systemic Strain Analysis}: Pressures on conference infrastructure were quantified using official statistics. The analysis included: (a) applying regression modeling to submission, acceptance, and rejection trends to measure review strain; (b) analyzing the research lifecycle lag against the conference calendar; and (c) forecasting attendance growth against venue capacities.
\end{itemize}

\paragraph{Metrics and Analysis Framework} We define and analyze key metrics to quantify the crisis.
\begin{itemize}
    \item \textbf{Research Pressure Metrics}: To measure the escalating pressure on the academic community, we define:
    \begin{itemize}
        \item \textit{Effective Faculties}: Adapted from \cite{iclrpoint}, this metric counts faculty in a given area, prorated by their cross-area affiliations (a faculty in $n$ areas contributes \$1/n\$ to each). This accounts for interdisciplinary researchers.
        \item \textit{Contribution per Person}: Calculated as $\textit{pub\_count} / \textit{effective\_faculties}$, this metric quantifies the average publication output per faculty member, revealing trends in academic productivity pressure.
    \end{itemize}
    \item \textbf{Environmental Impact}: We estimate carbon emissions using the standard formula: \textit{Activity Data $\times$ Emission Factor = Emissions}. Activity data includes round-trip flight distances from author institutions to conference venues and hotel stays. Emission factors are sourced from established calculators and environmental reports \cite{Gokus2019climate,vancouveremission}. Details are in Appendix \ref{sec:emission_factors}.
    \item \textbf{Research Lifecycle Lag}: We conceptually analyze the growing gap between a paper's initial public availability (e.g., on arXiv) and its formal presentation. While not quantified with a single metric here, this lag is a key indicator of inefficient knowledge dissemination.
    \item \textbf{Community Well-Being}: We employ sentiment analysis on Reddit threads, categorizing posts as ``Negative'', ``Mixed/Neutral'' and identifying the prevalence of explicit mental health keywords (e.g., ``anxiety'', ``burnout'', ``stress'').
\end{itemize}

\section{Challenges in Current AI Conferences}
\label{sec:badissue}

\begin{figure}[h]
    \centering
    \includegraphics[width=0.65\linewidth]{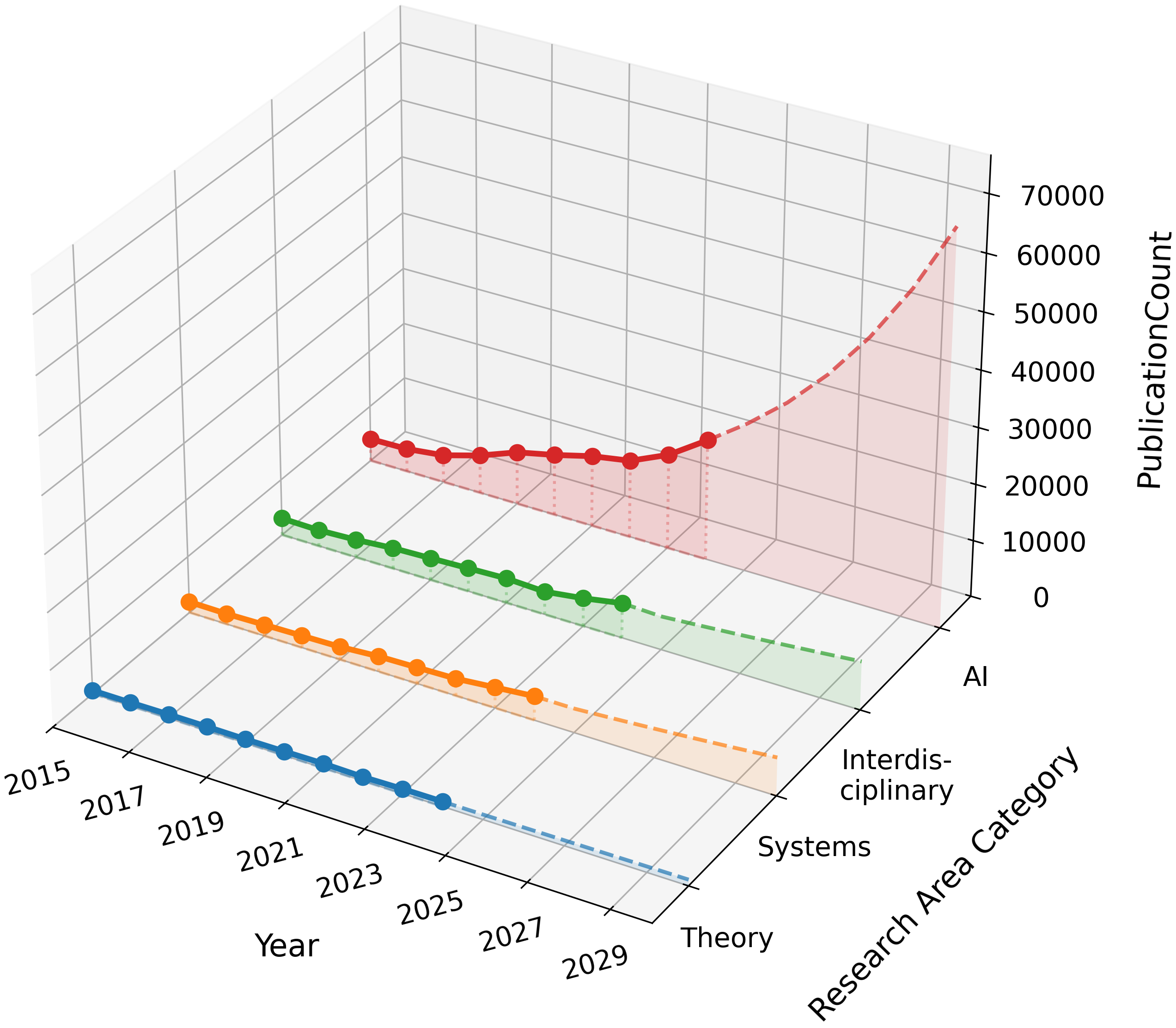}
    % \caption{Publication Growth: Ten-Year Historical Data and Ten-Year Future Estimate.}
    \caption{Publication growth in AI vs. Other CS Fields over the past decade, with 2030 projections.}
    \label{fig:pubcount}
\end{figure}

While the core goals of AI conferences remain essential, the current model increasingly deviates from these objectives due to unsustainable growth in submission volumes. This explosive expansion not only strains resources but also undermines the missions outlined in Section \ref{sec:goals}, transforming conferences from focused knowledge hubs into overwhelming, high-volume events that prioritize quantity over depth (Mamba \cite{gu2024mamba}'s rejection as a case).
To illustrate, we analyze publication trends across computer science subfields using conferences categorization (see Table in the appendix) and conference data in CSRankings \cite{csranking_github}. As shown in Figure \ref{fig:pubcount}, while fields like Theory, Systems, and Interdisciplinary areas exhibit relatively stable growth over the past decade, AI publications have experienced an almost exponential surge ($R^2 = 0.979, p<0.001$ under a log-linear regression model) over the past decade (especially true for 2023 and 2024, likely driven by the emergence and proliferation of LLMs). This trend is estimated to increase to over 65,000 (95\% prediction lower bound is 66,572) by 2030, more than tripling the 2024 figure of 21,400. 
Such unchecked growth has sparked widespread concerns such as the integrity of peer-review systems \cite{tran2020open,yang2025position,kim2025position}. This hyper-competitive environment, where the sheer volume of submissions overwhelms the review process, may also lower the barrier for academic misconduct, such as abuse of LLM writing \cite{oliveira2025humanaicollaborationacademicmisconduct} and LLM-based prompt injection \cite{promptinjection} in paper submissions, or make it harder to detect. Moreover, the annual cycle of conference submissions struggles to keep pace with the \textbf{shrinking lifecycle of AI research}. It is now a common phenomenon for work to be outdated by the time it is presented, and an initial rejection often leaves a paper with insufficient novelty for subsequent acceptance.

This surge in paper volumes also exacerbates a cascade of interconnected issues for authors and organizers as discussed below.

\subsection{“Publish or Perish” Treadmill: Hyper-Productivity}

The booming of AI domain submission signals a potential collapse in individual researchers' sustainable efforts, as the per-person output in AI becomes increasingly untenable.

\begin{figure}[H]
    \centering
    \includegraphics[width=0.75\linewidth]{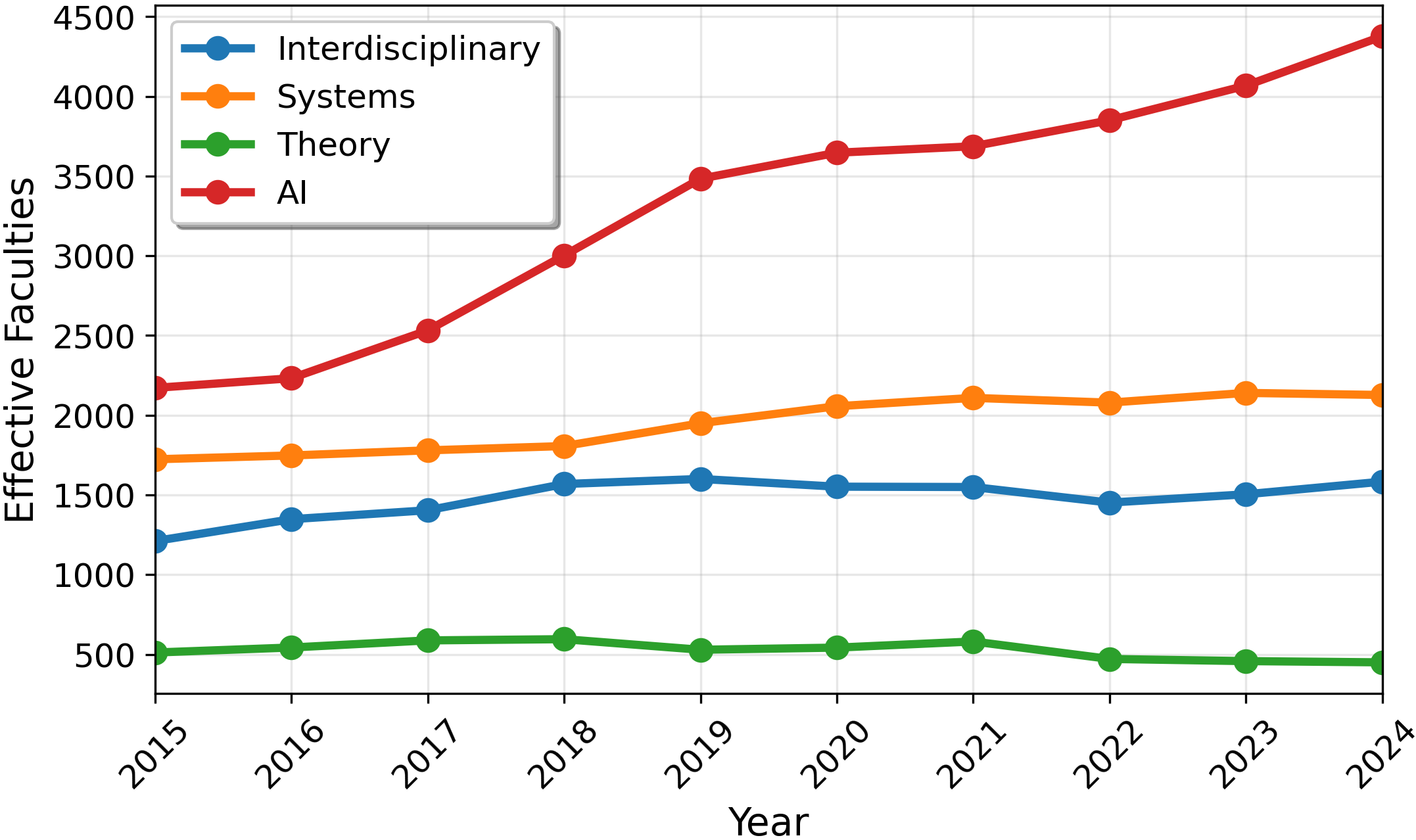}

    \caption{Growth of effective faculties by CS domain.}
    \label{fig:effective_faculties}
\end{figure}

To assess AI's broader impact on other fields, we leverage CSRankings data and adapt the concept of \textit{effective\_faculties} from \cite{iclrpoint}, which account for faculty members in each area (detailed in the appendix), prorated for cross-area affiliations (e.g., a faculty spanning n areas contributes 1/n to each), and plot yearly faculty trends for four key computer science domains in the appendix. Over the last decade, non-AI fields like Theory, Systems, and Interdisciplinary Applications have maintained relatively stable and lower faculty counts compared to AI, which has nearly doubled in size. More strikingly, from 2023 to 2024, Systems and Theory exhibit downward trends, potentially indicating faculty shifts toward multi-disciplinary pursuits. This may suggest a \textbf{``siphon effect''} from AI's rapid growth, potentially hindering the balanced development of other areas.

\begin{figure}[H]
    \centering
    \includegraphics[width=0.75\linewidth]{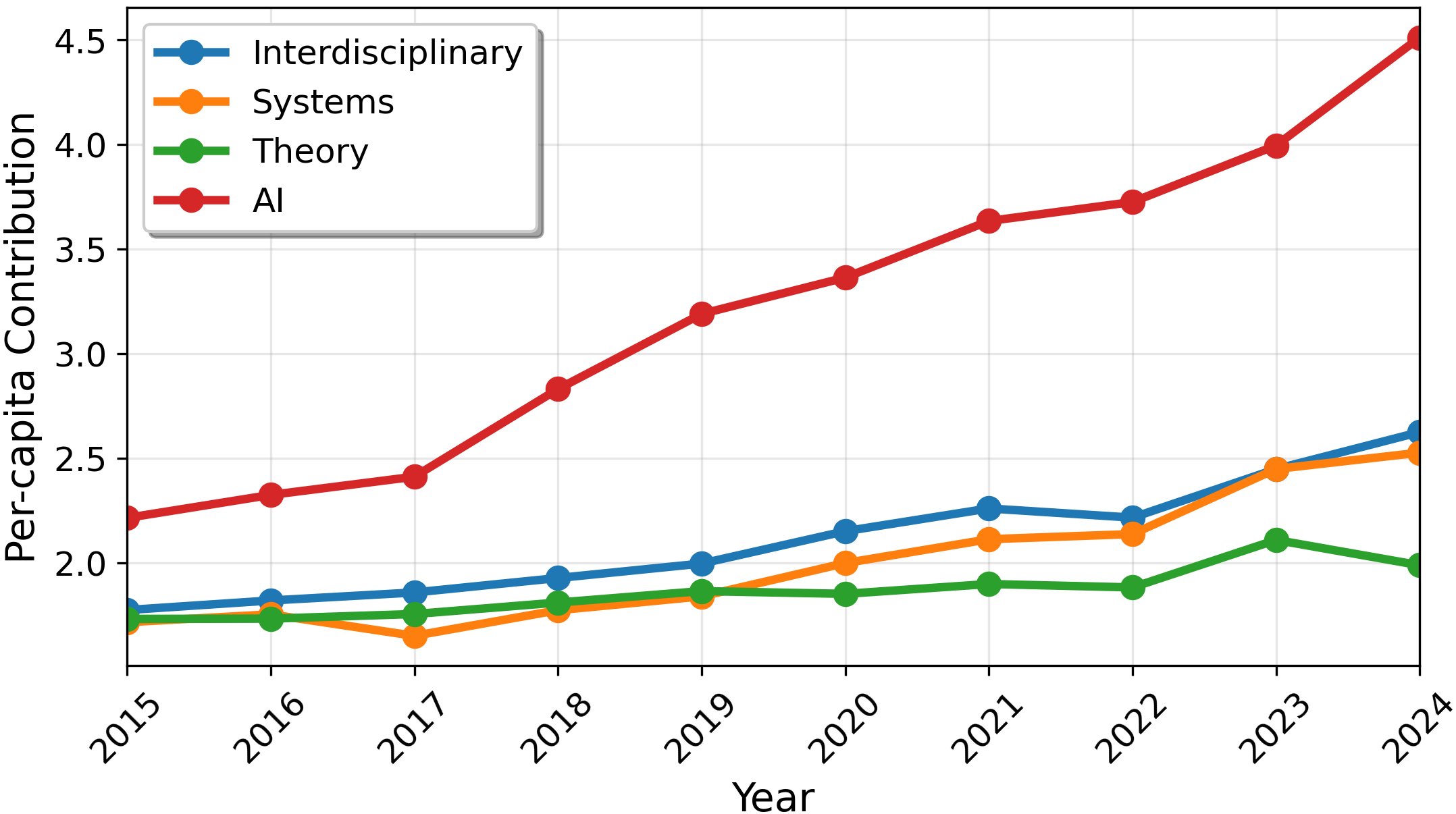}
    \caption{Average contribution per effective faculty.}
    \label{fig:contribution}
\end{figure}

What further complicates the situation is that publication counts are not merely increasing linearly with faculty growth; they are, in fact, also growing exponentially ($R^2=0.978, p<0.001$ under a log-linear regression model). We quantify per-capita contributions across fields through
\vspace{-0.2mm}
\begin{equation}
\mathrm{Per\text{-}capita\ Contribution} = \frac{\textit{Publication Count}}{\textit{Effective Faculties}}
\end{equation}

where \textit{pub\_count} represents the number of publications. As shown in Figure \ref{fig:contribution}, in the past decade, the exponential AI faculty publication rates have far outpaced the relative stability in other domains, reaching over twice the per-person output of non-AI fields and doubling within a decade to exceed 4.5 papers per person annually. If the exponential trend continues, output per faculty is \textbf{projected to exceed one paper every two months} (6 papers annually) \textbf{by 2030} (per-capita contribution 95\% prediction interval [6.34, 8.30]). This misalignment with AI conferences' core missions, particularly knowledge dissemination (as outlined in Section \ref{sec:goals}), diminishes meaningful exchanges by fostering a high-pressure environment that heightens mental health strains, such as anxiety impeding open collaborations and excessive competition stifling creativity and risk-taking issues we analyze further below.

\subsection{Environmental Toll: Mounting Carbon Footprint}

\begin{figure}[H]
    \centering
    \includegraphics[width=0.95\linewidth]{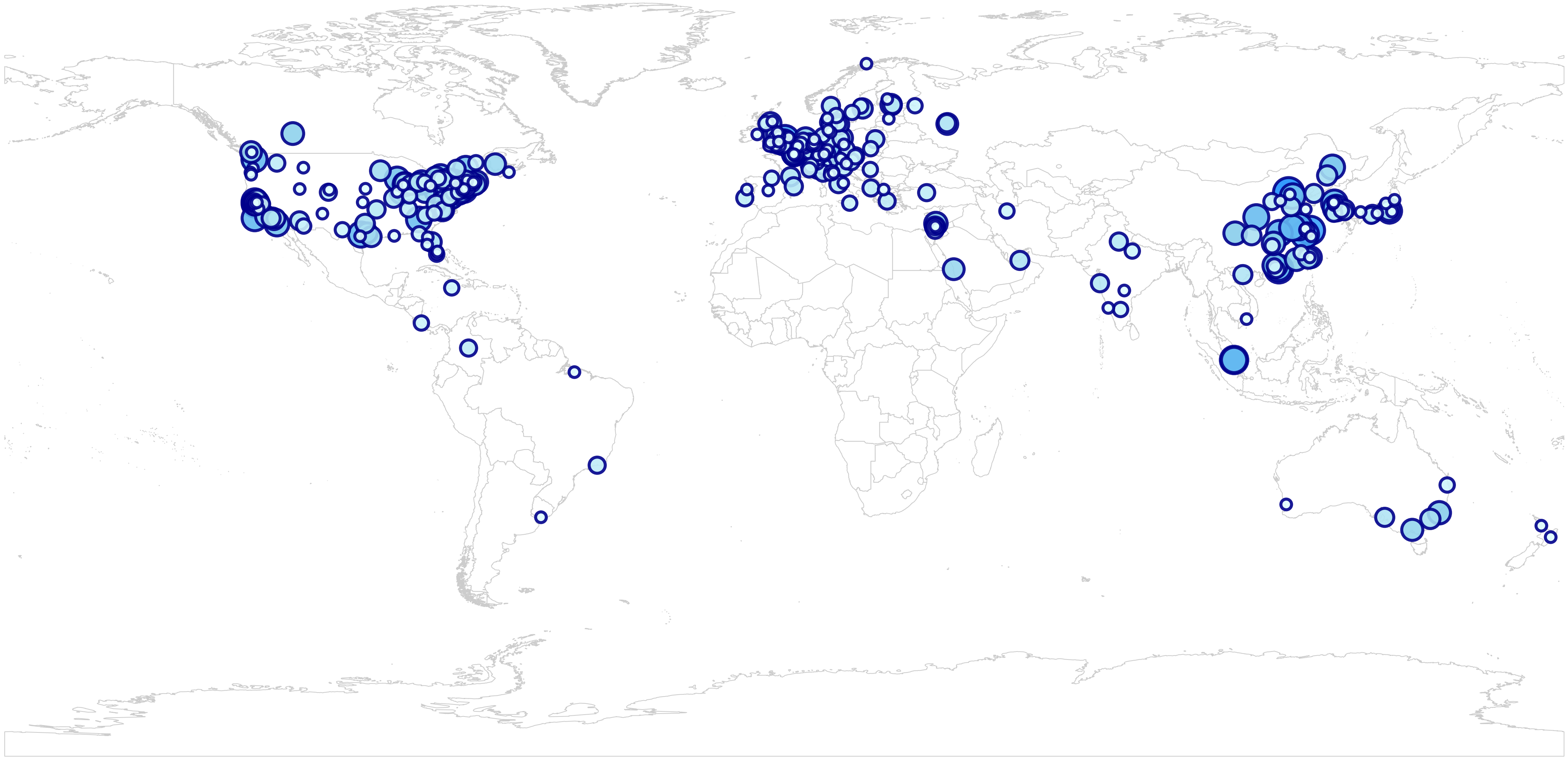}
    % \caption{NeurIPS 2024. more passenger, cycle larger, color buler. most of the cities (school) in the world have a flight.}
\caption{Geographic distribution of NeurIPS 2024 first authors.  Larger and darker blue circles represent a higher concentration.}
    \label{fig:worldmap}
\end{figure}

This surge in paper volumes also exacerbates a cascade of interconnected issues for authors, reviewers, and organizers, including elevated carbon footprints from increased travel and computing demands. Since most AI conferences \cite{neurips,iclr,icml} mandate in-person attendance for accepted authors, publication growth generates massive travel volumes. For the largest AI conference NeurIPS 2024, we estimate emissions for 3,836 unique first authors (from 4,037 before de-duplication, with 13,307 actual attendees \cite{nips_factsheet}) flying round-trip from their organizations, concentrated in Asia, Europe, and the Americas (as highlighted in Figure \ref{fig:worldmap}), to the Vancouver Convention Centre in Canada. Using the formula Activity data $\times$ Emission factors = emissions (details in the appendix), flight emissions alone reach 8,254 tCO$_2$e, exceeding the daily carbon output of Vancouver's approximately 680 thousand residents (based on 2.5 MtCO$_2$e annually \cite{vancouveremission}, or roughly 6,849 tCO$_2$e per day). These environmental impacts can challenge conferences' DEI initiatives, making it harder for researchers from underrepresented regions or those who are eco-conscious to participate. They also place unnecessary financial and time burdens on authors, including costs for visas and airfare.

\begin{figure}[t]
    \centering
    \includegraphics[width=0.75\linewidth]{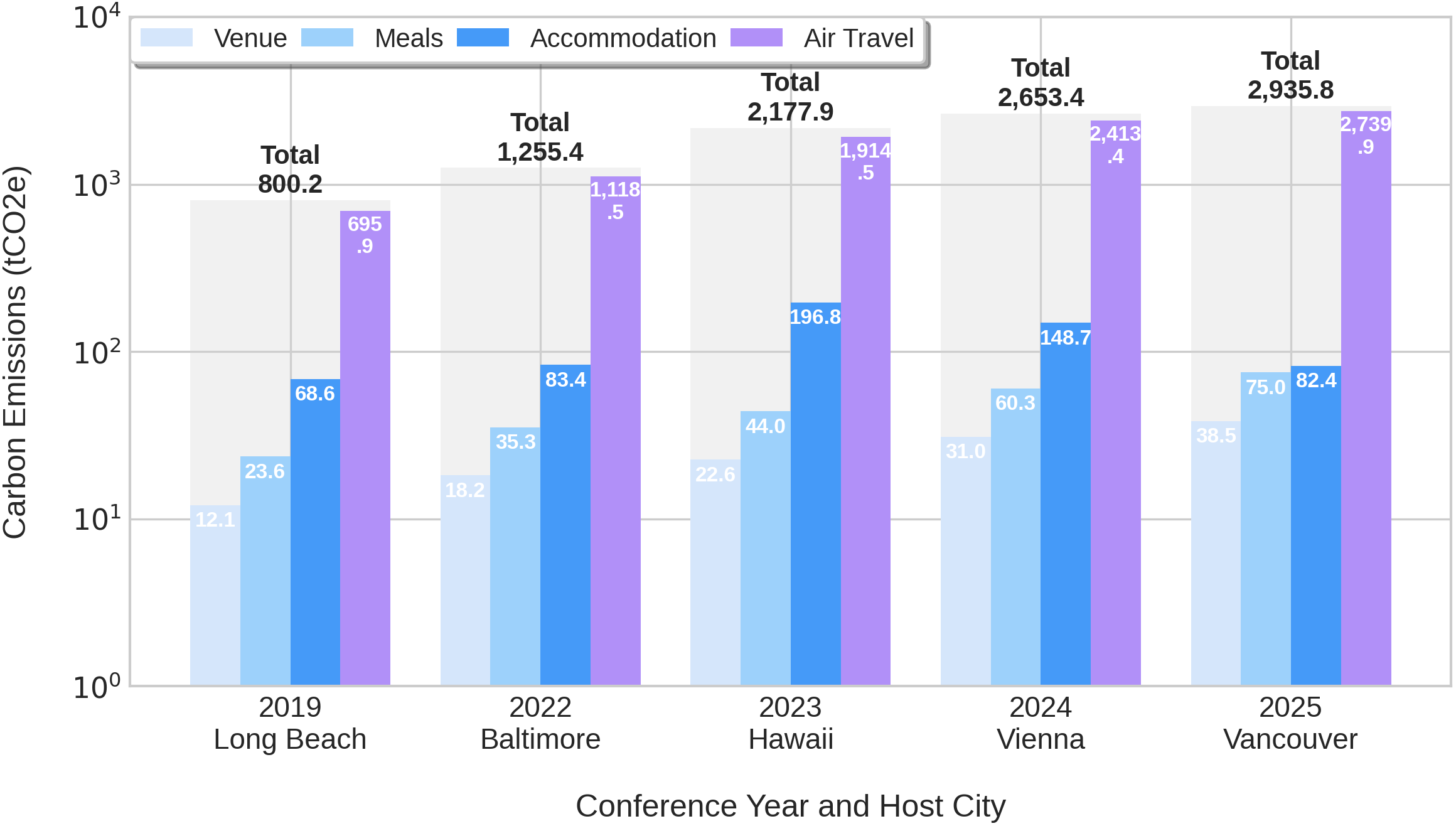}
    \caption{Surging carbon emissions from ICML.}
    \label{fig:carbon_pair}
\end{figure}

Prior studies have roughly examined travel impacts in non-computing conferences \cite{ajufo2021footprintcalculator,Gokus2019climate}, and provided services for carbon footprint calculations \cite{Jackle2022carbon,Gokus2019climate}. To extend the analysis to AI conference, we evaluate recent emissions from ICML (a 5-day event, based on top 200 first-author organizations), focusing on travel (transportation and accommodation) and venue (food and energy consumption); ICLR details appear in the appendix. As shown in Figure \ref{fig:carbon_pair}, transportation dominates emissions for non-hybrid ICML and ICLR, aligning with proportions from \cite{meetgreen}, while emissions from accepted authors have exploded, more than quadrupling over the past five years.

This trajectory tests organizers' sustainability commitments, rendering conferences environmentally untenable as attendance grows and violating the social contract. Moreover, economic pressures from accommodation and food as complaint for authors and venue selection factors for organizers will be explored further below.

\subsection{Human Cost: A Community Under Strain}

The growing scale and competitive nature of AI conferences also affect the emotional well-being and sentiment of different participant groups (authors, reviewers, attendees). For example, \cite{nicholls2022wellbeing} analyzed the academic impact on high pressure and job insecurity, and \cite{asa2018reflexivity} raised that doctoral students commonly experience identity anxiety and a lack of academic belonging. 
To systematically investigate these sentiments within the community, we analyzed discussions on public forums. We queried the Reddit subreddit, r/MachineLearning, using terms such as ICLR, NeurIPS, ICML, and “machine learning conferences” to identify the top 25 of the most relevant and high-activity (labelled ``hot'') discussion threads. We analyzed comments of each thread using VADER (Valence Aware Dictionary and sEntiment Reasoner), a lexicon-based sentiment analysis tool optimized for social media text. We focused on detecting indicators of negative sentiment or suggestions for improvement, and the sampled representative comments were subsequently aggregated to construct a sentence cloud in Figure \ref{fig:nips_venue}, highlighting recurring themes and concerns within the community and the high overlap with bad issues in this paper (see supplementary material for the sources).

\begin{center}
\begin{minipage}{0.65\linewidth}
\small
\textbf{Chart 1: Sentiment of Conference Threads (n=405)} \\
\textit{Proportion of threads with a negative sentiment.}

\texttt{Negative (289)} \hfill \textcolor{sentimentNegative}{\rule{5.5cm}{2.5ex}} \textbf{71.3\%} \\
\texttt{Mixed/Neutral (116)} \hfill \textcolor{sentimentNeutral}{\rule{2.2cm}{2.5ex}} \textbf{28.7\%}

\textbf{Chart 2: Negative Threads Analysis (n=289)} \\
\textit{Proportion with mental health keywords.}

\texttt{Contains MH (100)} \hfill \textcolor{mhKeywords}{\rule{3.0cm}{2.5ex}} \textbf{34.6\%} \\
\texttt{No MH (189)} \hfill \textcolor{noMhKeywords}{\rule{6.0cm}{2.5ex}} \textbf{65.4\%}
\end{minipage}
\end{center}
Searching Reddit for keywords such as ``ICLR'', ``ICML'', ``NeurIPS'', and ``Machine Learning'' reveals consistent community-level negativity toward top-tier ML conferences. Explicit emotional language (e.g., anxiety, frustrated, distressed, disgusted) frequently appears, with some posts even tagged under mental health. Among the top ten relevant posts, several contain comments suggesting deteriorating mental well-being (see Table in the appendix). We further conduct a large-scale visual analysis of 405 related threads from the \texttt{r/MachineLearning} subreddit (2022–2025).

As analyzed in Charts 1\&2, over 71\% of conference-related discussions express negative sentiment, indicating widespread community dissatisfaction. More than a third (34.6\%) of these negative threads mention terms related to mental health distress, such as `anxiety,' `burnout,' and `stress.'

This toxic environment, fueled by public criticism and pressure, undermines the conference's core goals. It erodes community building by replacing collaboration with anxiety and stifles genuine knowledge sharing by discouraging risk-taking essential for breakthroughs. It also violates the Social Contract of DEI, as a culture of anxiety is inherently not inclusive. This psychological strain is a symptom of the system’s overwhelming scale.

Authors' psychological stress may also stem from related dynamics, as evidenced by NeurIPS submission statistics in Figure \ref{fig:nips_venue}. Under the pressure of massive submissions, accepted papers exhibit roughly linear growth ($R^2 = 0.964, p<0.001$ under a linear regression model,), while rejections surge at a much faster, near-exponential rate ($R^2 = 0.937$, $p<0.001$ under a log-linear regression model). This inevitably drives more resubmissions, further straining reviewer capacity. Compounding the issue, state-of-the-art (SOTA) benchmarks from prior submissions may become outdated by the next cycle, fostering a distorted mindset among authors who chase overly positive reviews at the expense of genuine innovation. For instance, studies suggest that AI agent capabilities double approximately every seven months \cite{kwa20257month, ord2025there}. Since the conference cycle from submission to presentation also lasts nearly seven months, this means \textbf{research can be outdated by the time it is published}, rendering a significant portion of the community’s effort inefficient. Such dynamics undermine community building and knowledge sharing. Additionally, the sheer volume of acceptances burdens organizers, with exploding attendance, visa restrictions, and other logistical constraints prompting NeurIPS to adopt a hybrid online-offline format since 2022. Pushing this concept further, NeurIPS 2025 will host parallel main conferences in Mexico City \cite{neurips2025_mexico} and Copenhagen \cite{neurips2025_copenhagen}. These venue-related challenges are discussed below.

\subsection{Physical Breaking Point: Venue Overload}
\begin{figure}[t]
    \centering
    \begin{minipage}{0.47\linewidth}
        \centering
        \includegraphics[width=\linewidth]{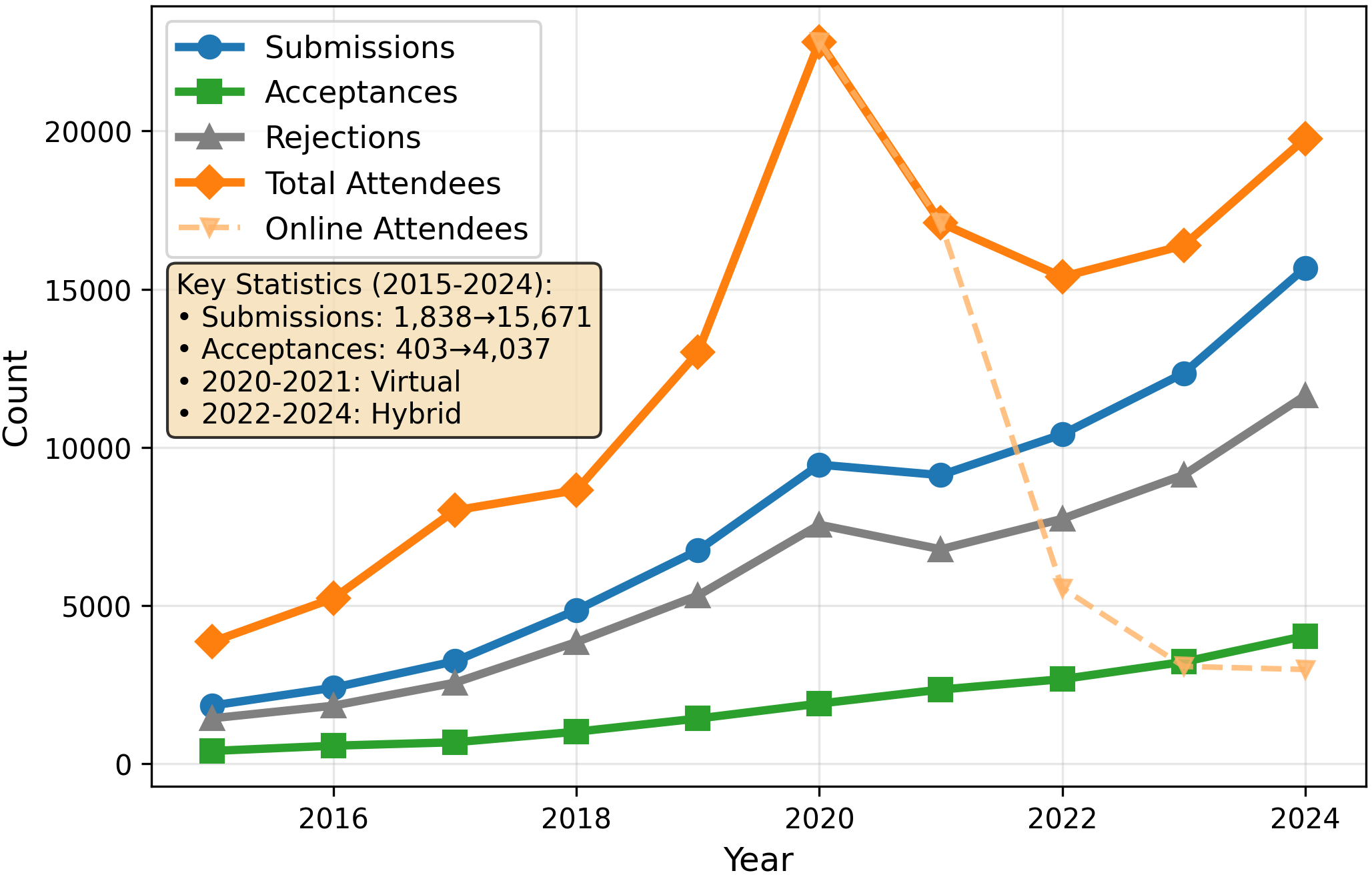}
        \caption{Trends of submissions, acceptances, rejections and attendees at NeurIPS.}
        \label{fig:nips_venue}
    \end{minipage}%
    \hfill
    \begin{minipage}{0.45\linewidth}
        \centering
        \includegraphics[width=\linewidth]{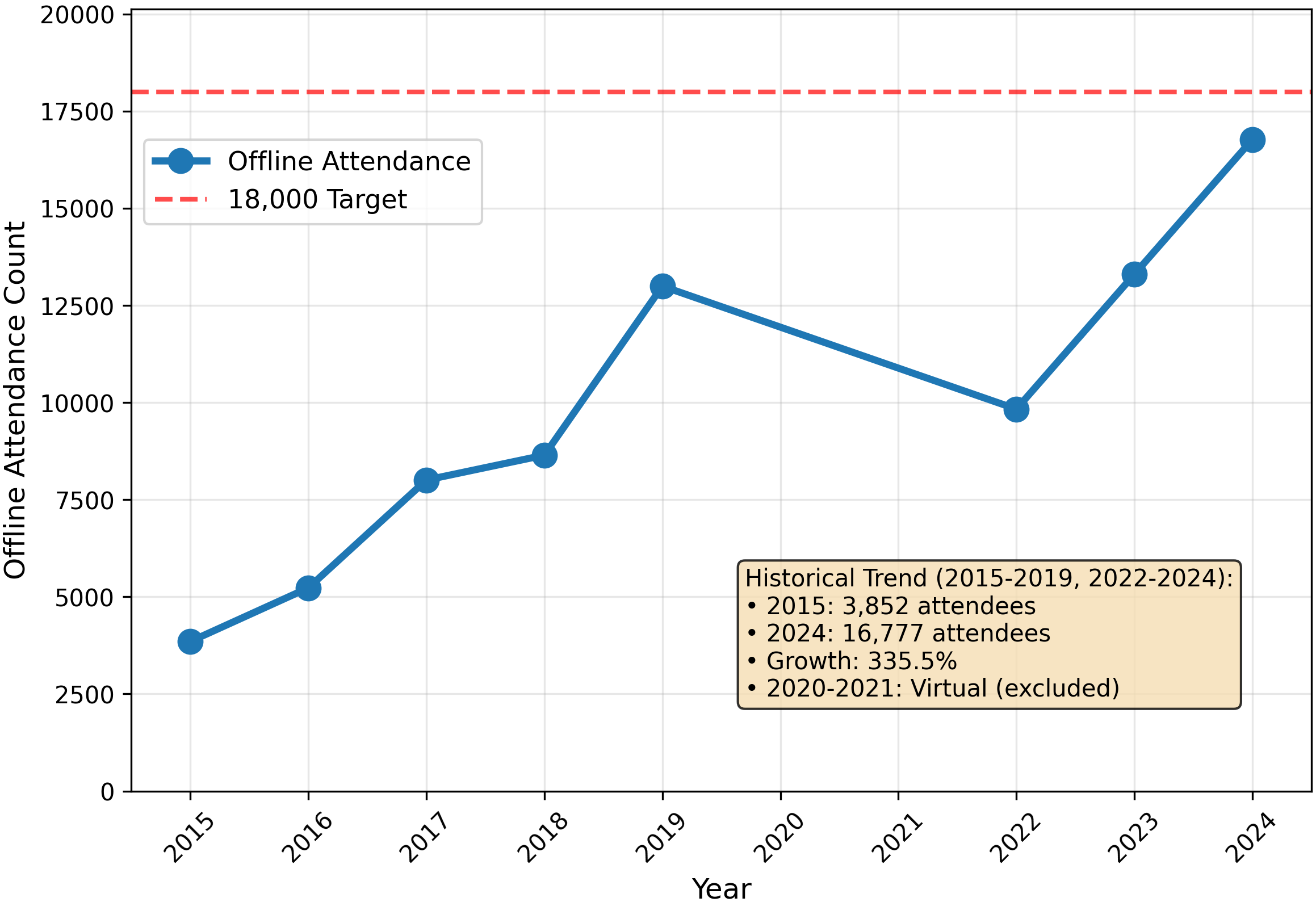}
        \caption{Trend of in-person attendance at NeurIPS over recent years.}
        \label{fig:nips_venue_predict}
    \end{minipage}
\end{figure}

As academic conferences grow in scale and scope, physical venues are increasingly unable to keep pace. This pressure is particularly evident at flagship AI conferences such as NeurIPS. The Vancouver Convention Centre, host of NeurIPS 2024, has a maximum capacity of approximately 18,000 attendees \cite{vcc_about}. Recognizing that registration demand was nearing this limit (as reflected in Figure \ref{fig:nips_venue_predict}), NeurIPS 2024 implemented a lottery system \cite{nips_lottery} for non-author registrations. This logistical constraint has fueled community concerns that venue limits could pressure organizers to force a lower acceptance rate or even push to reject already accepted papers \cite{nips25rejectduetovenue}. While necessary from a logistical perspective, this policy introduced a layer of artificial scarcity that limits participation for students, early-career researchers, and unaffiliated attendees who may benefit most from in-person engagement.

The effects of such constraints go beyond logistics. Reduced access blocks opportunities for spontaneous interaction, mentorship, and community building, particularly for those outside established research networks. It also compromises the principle of equitable participation that academic conferences are meant to uphold. As physical attendance continues to grow, even the largest venues are unlikely to provide an inclusive solution, revealing structural limits of the centralized model and highlighting the need for a more flexible and equitable alternative.

\section{Evaluating Incremental Fixes: Why Palliative Measures Fail}
\label{sec:fix}

In response to the growing challenges detailed in Section \ref{sec:badissue}, the AI community has begun to experiment with incremental adjustments to the traditional conference model. However, these well-intentioned measures tend to address the symptoms rather than the root causes of the system’s unsustainable, centralized growth. As a result, core issues such as overwhelming submission volumes, environmental impact, and barriers to equity and inclusion remain largely unresolved.

A case in point is the proposal to limit submissions per author \cite{kdd} applied to manage submission volume. However, this supply-side constraint is the zero-sum game of submission caps. It does not reduce the immense institutional pressure to publish; it merely shifts it, forcing researchers to be more strategic but no less stressed. Such caps can disproportionately penalize junior researchers who need to build their publication record, or productive labs working on multiple innovative fronts. Most importantly, it does nothing to alleviate the systemic ``publish-or-perish'' culture that drives burnout and discourages high-risk, long-term research. It’s an administrative patch on a cultural wound.

Similarly, the adoption of multi-site or satellite conferences, such as NeurIPS 2025, hosting parallel events in Mexico City and Copenhagen \cite{neurips2025_mexico, neurips2025_copenhagen}, is a direct response to venue capacity limits and aims to reduce long-haul travel for some participants. However, a multi-site conference still concentrates the entire community’s review burden and mental health anxiety into the same frantic, annual cycles. It maintains the centralized authority and the high-stakes, all-or-nothing evaluation process. It mitigates some travel but still fails to address the core issues of reviewer burnout, escalating author effort, and the exclusion inherent in any time-bound, high-cost event. It may also inadvertently create a two-tiered system, where one venue is perceived as more prestigious, thus contradicting the goal of equitable participation under the social contract.

Incremental fixes to the centralized model are no longer sufficient. A lasting solution requires dismantling its core components, including centralized structure, synchronized deadlines, and a monolithic format, and rebuilding the system around decentralization, flexibility, and community agency. Having established the limitations of incremental reform, we now introduce the Community-Federated Conference model.

\section{A Path Forward: Community-Federated Conferences}

Our preceding analysis reveals that the traditional centralized, single-venue conference model is collapsing under its own weight. It imposes escalating environmental, financial, and psychological costs that increasingly undermine the core mission of scholarly assembly. Existing responses, such as hybrid formats or multi-site replications (e.g., IJCAI 2025 or NeurIPS 2025), offer only incremental and temporary relief. These approaches replicate existing structures at smaller scales without confronting the systemic roots of the problem. We therefore propose a fundamental rethinking: the Community-Federated Conference (CFC) model.

The CFC model presents a sustainable, equitable, and scalable framework for organizing academic conferences. Its guiding principle can be summarized as ``Global Standards, Local Realization'', which is achieved by decoupling the three traditional functions of conferences: (1) peer review and publication, (2) knowledge dissemination, and (3) community building. These functions are restructured into distinct but interconnected layers.

\paragraph{Layer 1: Unified Global Peer Review and Publication.}
This layer introduces a centralized, high-quality digital platform managed by a consortium of academic societies (e.g., AAAI, ACM). Submissions and reviews occur on a rolling basis throughout the year, independent of any physical meeting. This temporal decoupling eases reviewer burden and allows for more thoughtful feedback, addressing concerns raised in Section~\ref{sec:badissue}. Accepted papers are published in globally recognized proceedings, ensuring academic credit and visibility. This process can also be supported by automation. NLP-based tools can assist in reviewer-paper matching, detect conflicts of interest, and flag anomalous reviews, reducing the manual workload involved in managing large-scale conferences.

\paragraph{Layer 2: Federated Regional Hubs for Dissemination and Networking.}
Once accepted, authors present their work at a regional hub of their choice. These hubs, organized by universities, local research labs, or student-led groups, typically host 500 to 1,500 participants. This federated model directly addresses the major logistical and sustainability challenges facing today's conferences. It eliminates the need for mega-venues, reduces carbon emissions by encouraging regional travel, and lowers financial barriers, promoting greater diversity, equity, and inclusion. In these smaller and more focused gatherings, researchers can engage in meaningful interactions, avoiding the anonymity and psychological strain of mega-conferences, as discussed in Section~\ref{sec:badissue}.

\paragraph{Layer 3: Digital Synchronization and Collaboration.}
What sets the CFC model apart from a collection of independent events is a unifying digital layer. This includes a Global Plenary track that live-streams keynotes and award talks from a rotating anchor hub to all other hubs. In addition, permanent digital poster halls allow for asynchronous discussion of all accepted papers, while thematic virtual channels (e.g., Slack or Discord) connect researchers working on similar topics across regions. Through this structure, local participation remains deeply connected to global discourse, enabling broad collaboration without the need for physical convergence.

Unlike centrally managed multi-site formats, the CFC model is built from the ground up by the community. And unlike traditional hybrid meetings that often treat remote participation as secondary, the digital layer in CFC is fully integrated and equally valued. By disentangling and distributing the core functions of academic conferences, the CFC model offers a resilient and forward-looking architecture. It not only addresses the shortcomings of the current system but also advances the core values of inclusivity, sustainability, and intellectual exchange.

\section{Conclusion}
The prevailing centralized model for AI conferences is becoming increasingly unsustainable. It imposes significant environmental costs, contributes to mental health challenges, and reinforces inequitable access. Incremental adjustments have failed to resolve the systemic pressures arising from hyper-competition, reviewer overload, and logistical constraints. We propose a fundamental shift to the Community-Federated Conference (CFC) model, which separates peer review, presentation, and community interaction into globally coordinated but locally executed components. It supports long-term sustainability, improves accessibility, and fosters deeper academic engagement. 

Preserving the vitality and integrity of the AI research ecosystem requires bold, structural reform. We invite the community to reimagine how we gather, share knowledge, and build scientific community. The code and accompanying materials are available at: \url{https://github.com/NuoJohnChen/AI_Conf_Crisis}.

\section*{Acknowledgments}
We thank Bryan Hooi, Zhaomin Wu and Chunxian Zhang for their helpful feedback.

% \nocite{conv00}
\clearpage
\begin{hyphenrules}{nohyphenation}
\setlength{\bibsep}{.5ex plus .8ex}
\bibliographystyle{unsrtnat}
\bibliography{main}
\end{hyphenrules}

\clearpage
\appendix

\section{Detailed Emission Factors}
\label{sec:emission_factors}

Table \ref{tab:emission_factors} shows the emission factors for AI conference components.

\begin{table*}[h]
\centering
\caption{Emission Factors for AI Conference Components.}
\label{tab:emission_factors}
\resizebox{\textwidth}{!}{
\setlength{\tabcolsep}{3pt}
\begin{tabular}{l l l}
\hline
\textbf{Emission Category} & \textbf{Selected Factor (Value and Unit)} & \textbf{Primary Source(s)} \\
\hline
Air Travel & 0.158 kg CO$_2$e / passenger.km & \cite{airtravelcarbon} \\
Accommodation (Kigali, 2023) & 28.0 kg CO$_2$e / room-night & \cite{rwandaGEF}, \cite{defraHospitality} \\
Accommodation (Vienna, 2024) & 13.9 kg CO$_2$e / room-night & \cite{ukBEIS}, \cite{defraAustria} \\
Accommodation (Singapore, 2025) & 24.5 kg CO$_2$e / room-night & \cite{ukBEISDEFRA} \\
Catering (Meals) & 1.88 kg CO$_2$e / meal & \cite{scarborough2014}, \cite{imperialGrantham} \\
Venue Operations & 2.896 kg CO$_2$e / person/day & \cite{babka2024} \\
\hline
\end{tabular}}
\end{table*}

The data in table \ref{tab:emission_factors} is retrieved from the UK's Department for Environment, Food and Rural Affairs (DEFRA) GreenHouse Gas (GHG) Conversion Factors 2025\cite{gasreport2025department}.

\section{Detailed AI Conference Categories}
\label{sec:AIconfcate}
Table \ref{tab:cs_conferences} shows computer science areas and their corresponding top-tier conferences.
\begin{table*}
\centering
\caption{Computer science areas and their corresponding top-tier conferences. The list of conferences is based on the CSRankings website.}
\label{tab:cs_conferences}
\resizebox{\textwidth}{!}{
\setlength{\tabcolsep}{3pt}
\begin{tabular}{llll}
\toprule
\textbf{Parent Area} & \textbf{Abbreviation} & \textbf{Area} & \textbf{Conferences} \\
\midrule
\multirow{5}{*}{AI} & ai & Artificial intelligence & AAAI, IJCAI \\
 & vision & Computer vision & CVPR, ECCV, ICCV \\
 & mlmining & Machine learning & ICLR, ICML, NeurIPS \\
 & nlp & Natural language processing & ACL, EMNLP, NAACL \\
 & inforet & The Web \& information retrieval & SIGIR, WWW \\
\midrule
\multirow{12}{*}{Systems} & arch & Computer architecture & ASPLOS, ISCA, MICRO \\
 & comm & Computer networks & SIGCOMM, NSDI \\
 & sec & Computer security & CCS, IEEE S\&P (``Oakland''), USENIX Security \\
 & mod & Databases & SIGMOD, VLDB \\
 & da & Design automation & DAC, ICCAD \\
 & bed & Embedded \& real-time systems & EMSOFT, RTAS, RTSS \\
 & hpc & High-performance computing & HPDC, ICS, SC \\
 & mobile & Mobile computing & MobiCom, MobiSys, SenSys \\
 & metrics & Measurement \& perf. analysis & IMC, SIGMETRICS \\
 & ops & Operating systems & OSDI, SOSP \\
 & plan & Programming languages & PLDI, POPL \\
 & soft & Software engineering & FSE, ICSE \\
\midrule
\multirow{3}{*}{Theory} & act & Algorithms \& complexity & FOCS, SODA, STOC \\
 & crypt & Cryptography & CRYPTO, EuroCrypt \\
 & log & Logic \& verification & CAV, LICS \\
\midrule
\multirow{7}{*}{Interdisciplinary Areas} & bio & Comp. bio \& bioinformatics & ISMB, RECOMB \\
 & graph & Computer graphics & SIGGRAPH, SIGGRAPH Asia \\
 & csed & Computer science education & SIGCSE \\
 & ecom & Economics \& computation & EC, WINE \\
 & chi & Human-computer interaction & CHI, UbiComp / Pervasive / IMWUT, UIST \\
 & robotics & Robotics & ICRA, IROS, RSS \\
 & visualization & Visualization & VIS, VR \\
\bottomrule
\end{tabular}}
\end{table*}

\section{Conference Attendance Data}
Table \ref{tab:conference_data} shows ICLR Conference Attendance Data (2023-2025).

\begin{table*}
\centering
\caption{ICLR Conference Attendance Data (2023-2025)}
\label{tab:conference_data}
\small % 使用小号字体以确保表格能放入页面
\resizebox{\textwidth}{!}{
\setlength{\tabcolsep}{3pt}
\begin{tabular}{c r l l r c c r p{5cm}}
\toprule
\textbf{Year} & \textbf{\makecell{Total \\ Participants}} & \textbf{\makecell{Origin \\ (First Author)}} & \textbf{Destination} & \textbf{\makecell{No. Attendees \\ (Est. Live)}} & \textbf{\makecell{Hotel \\ Nights}} & \textbf{Meals} & \textbf{\makecell{Air Distance \\ (KM)}} & \textbf{Remarks} \\
\midrule

% --- 2025 Data ---
\multirow{15}{*}{2025} & \multirow{15}{*}{11039} & United States & \multirow{15}{*}{Singapore} & 2172 & 5 & 15 & 14910 & \multirow{15}{=}{\parbox{5cm}{11,039 participants spanning 85 countries; 10,435 in-person. \newline\newline Only countries with over 100 participants are listed.}} \\
& & China, Peoples Republic of China & & 1946 & 5 & 15 & 4245 & \\
& & Singapore & & 947 & 0 & 15 & 0 & \\
& & United Kingdom & & 583 & 5 & 15 & 10859 & \\
& & Korea, Republic of Korea & & 528 & 5 & 15 & 4669 & \\
& & Germany & & 398 & 5 & 15 & 10311 & \\
& & Canada & & 308 & 5 & 15 & 13500 & \\
& & Switzerland & & 227 & 5 & 15 & 10300 & \\
& & Japan & & 213 & 5 & 15 & 5340 & \\
& & Australia & & 223 & 5 & 15 & 6293 & \\
& & Hong Kong SAR, China & & 207 & 5 & 15 & 2555 & \\
& & France & & 164 & 5 & 15 & 10724 & \\
& & Israel & & 137 & 5 & 15 & 8002 & \\
& & India & & 117 & 5 & 15 & 4150 & \\
& & Netherlands & & 106 & 5 & 15 & 10513 & \\
\midrule

% --- 2024 Data ---
\multirow{11}{*}{2024} & \multirow{11}{*}{6553} & United States & \multirow{11}{*}{Vienna, Austria} & 1647 & 5 & 15 & 6814 & \multirow{11}{=}{\parbox{5cm}{6,533 participants spanning 79 countries; 5,938 in-person. \newline\newline Only countries with over 100 participants are listed.}} \\
& & China, Peoples Republic of China & & 814 & 5 & 15 & 7496 & \\
& & Germany & & 494 & 5 & 15 & 539 & \\
& & United Kingdom & & 452 & 5 & 15 & 1277 & \\
& & Korea, Republic of Korea & & 315 & 5 & 15 & 8270 & \\
& & Canada & & 244 & 5 & 15 & 6950 & \\
& & Australia & & 215 & 5 & 15 & 15767 & \\
& & Switzerland & & 204 & 5 & 15 & 632 & \\
& & France & & 160 & 5 & 15 & 1038 & \\
& & Singapore & & 118 & 5 & 15 & 9698 & \\
& & Japan & & 104 & 5 & 15 & 9150 & \\
& & Netherlands & & 102 & 5 & 15 & 950 & \\
\midrule

% --- 2023 Data ---
\multirow{7}{*}{2023} & \multirow{7}{*}{3758} & United States & \multirow{7}{*}{Kigali, Rwanda} & 1960 & 5 & 15 & 11500 & \multirow{7}{=}{\parbox{5cm}{3,758 participants spanning 73 countries; 2,015 in-person. \newline\newline First in-person conference since Covid. \newline Only countries with over 100 participants are listed.}} \\
& & China, Peoples Republic of China & & 1222 & 5 & 15 & 10000 & \\
& & United Kingdom & & 273 & 5 & 15 & 6590 & \\
& & Korea, Republic of Korea & & 229 & 5 & 15 & 10760 & \\
& & Germany & & 176 & 5 & 15 & 6200 & \\
& & Canada & & 163 & 5 & 15 & 11400 & \\
& & Singapore & & 124 & 5 & 15 & 8208 & \\
& & Switzerland & & 102 & 5 & 15 & 5800 & \\

\bottomrule
\end{tabular}}
\end{table*}

\section{Carbon Emissions from ICLR.}
\begin{figure}[H]
    \centering
    \includegraphics[width=0.95\linewidth]{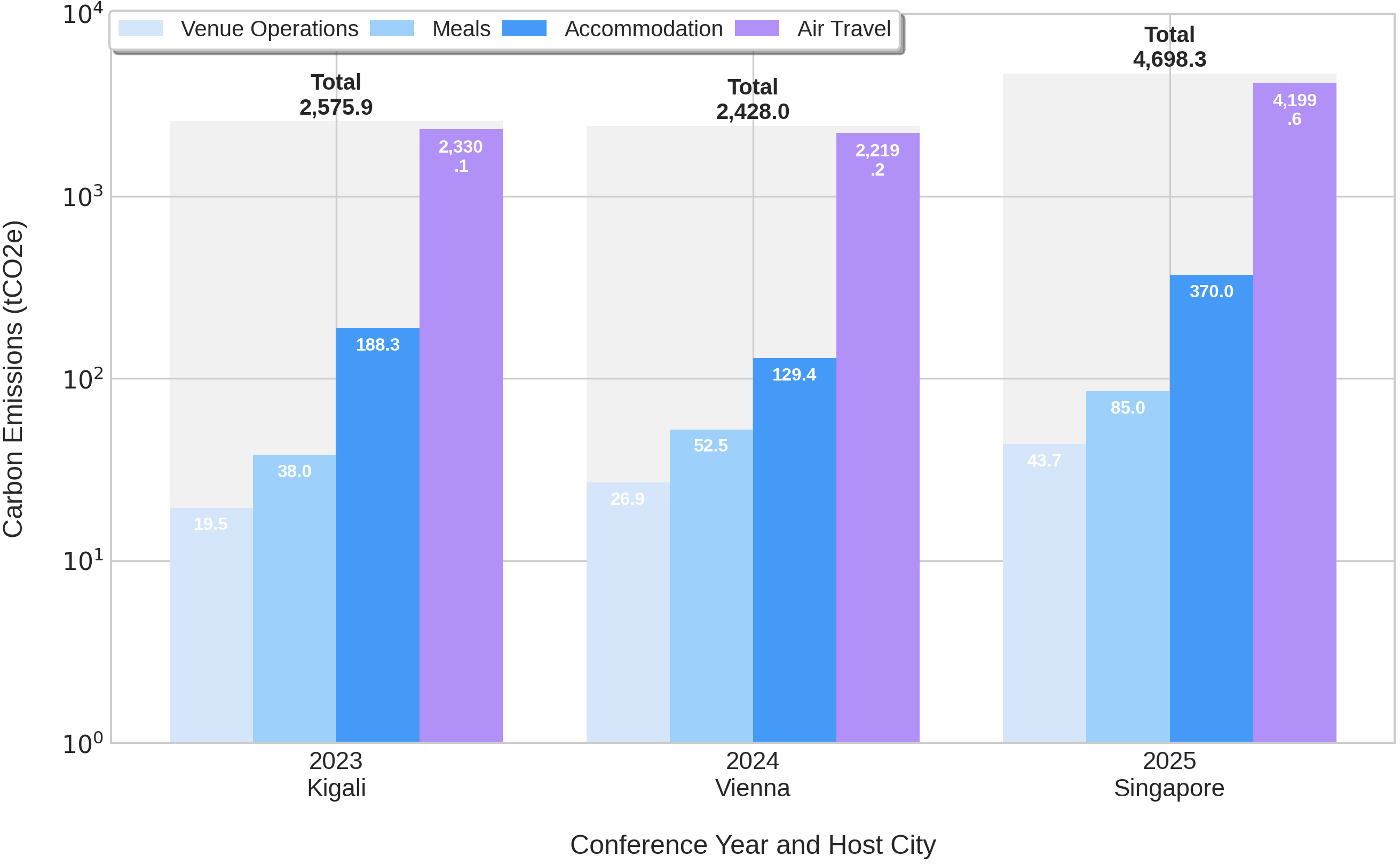}
    \caption{Carbon Emissions from ICLR.}
    \label{fig:carbon_iclr}
\end{figure}

Table \ref{fig:carbon_iclr} shows the carbon Emissions from ICLR.

\section{Reddit Discussions}
Table \ref{tab:iclr_mental_health} shows the examples of ICLR Reddit discussions.

% The source of the analyzed comments in the paper is available in \ref{https://docs.google.com/spreadsheets/d/1O1Vbbq6Liy2KGzLMOaziGkN5yioOlFVX/edit?gid=500222994#gid=500222994}.

\begin{table*}
\centering
\caption{Examples of ICLR Reddit Discussions Reflecting Mental Health Struggles.}
\label{tab:iclr_mental_health}
\resizebox{\textwidth}{!}{
\setlength{\tabcolsep}{3pt}
\begin{tabular}{@{}p{4.3cm}lllp{6.2cm}@{}}
\toprule
\textbf{Thread Title} & \textbf{Date} & \textbf{MH Tag} & \textbf{Sentiment} & \textbf{Key Quote} \\
\midrule
ICLR 2025 paper decisions. & Nov 2024 & Review fatigue & Negative & \textit{``it was exhausting... spent the entire discussion period addressing each new `problem'''} \\
{[D] ICLR 2025 Paper Reviews Discussion.} & Oct 2024 & Stress/anxiety & Negative & \textit{``Reviewers are clearly not... putting in the time... 30 minutes isn't enough''} \\
Quality of ICLR papers. & Nov 2024 & Burnout & Negative & \textit{``I reviewed... some of the worst papers... incremental... perverse incentives.''} \\
ICLR 2023 reviews are out. How was your experience? & May 2023 & Anxiety & Negative & \textit{``I didn't read them yet, because of anxiety.''} \\
ICLR 2024 decisions are coming out today. & Jan 2024 & Stress & Negative & \textit{``my paper... if this time does not work again, I will be very sad... research in ml is a pain.''} \\
Toxic reviews at ICLR? & Jun 2025 & Toxicity & Negative & \textit{collective frustrations (inferred)} \\
ICLR submissions should not be public. & Jun 2025 & Anxiety/stress & Negative & \textit{``I am absolutely disgusted by their academic integrity...''} \\
When are ICLR reviews out? & May 2018 & Anxiety & Negative & \textit{``Seen anxiety about waiting — historic relevance''} \\
Why is everybody surprised that Mamba got rejected for? & Feb 2024 & Frustration & Negative & \textit{``reviewers measure contribution... I don't think this warrants a rejection.''} \\
ICLR 2022 RESULTS ARE OUT & Jun 2022 & Anxiety & Negative & \textit{``Reviewer2 gave me a lot of anxiety about the scores...''} \\
\bottomrule
\end{tabular}
}
% \smallskip
% \footnotesize \textcolor{gray}{Note: Special tags like [D] preserved; Timespan shows recurring issues from 2018-2025.}
\end{table*}

\end{document}